\documentclass[preprint]{ptephy_v1}

\usepackage{hyperref}
\usepackage{bm}
\newcommand{\ri}{{\rm i}}

\begin{document}

\title{Kaluza-Klein Schwinger effect}


\author{Yusuke Yamada}
\affil{Waseda Institute for Advanced Study, Waseda University, 1-21-1 Nishi Waseda, Shinjuku, Tokyo 169-0051, Japan\email{y-yamada@aoni.waseda.jp}}


\begin{abstract}%
We show that electric fields in compactified spaces may produce Kaluza-Klein (KK) particles even when the energy of electric fields is smaller than KK scale. As an illustrating example, we consider a charged massless complex scalar coupled to U(1) gauge theory in $\mathbb{R}^{1,3}\times {\mathbb S}^1$ and discuss the effect of background gauge potential along a compact direction. The electric field produces the charged Kaluza-Klein particle non-perturbatively, which we call KK Schwinger effect. We quantitatively show that KK modes can be produced even when the electric field energy is much below the KK scale. The mechanism is rather general and similar phenomena would occur in any compactification models when a gauge potential along compact direction evolves in time and experiences large enough field excursion. We also discuss the subtlety of four dimensional effective theory truncated by KK modes at an initial time, when the electric field is turned on.
\end{abstract}


\maketitle
\section{Introduction}
\label{introduction}
The existence of extra spaces is one of the most important predictions of string theory. It is common to assume that the extra spaces are compactified such that the KK modes having momenta along compact directions become heavy enough. Truncating such heavy KK modes while leaving light or massless modes yields four dimensional (4D) effective theory. Thus derived effective theory is usually thought to be correct as long as the energy scale under consideration is smaller than the KK scale roughly given by the inverse of the compactification radius. 

Gauge theory in higher-dimensional theory contains gauge fields along compactified directions. In 4D effective theory, the zero modes of such components unless projected out e.g. by orbifold conditions are typically light as their masses are forbidden by the original gauge symmetry. Such light fields in 4D effective theory may play various important roles as dark matters or as inflaton fields for cosmic inflation in the early Universe. Indeed, the extra-natural inflation~\cite{Arkani-Hamed:2003xts,Arkani-Hamed:2003wrq} is the model where inflation is driven by the Wilson line phase being the extra-dimensional gauge potential integrated along compact spaces. From the original higher-dimensional theory viewpoint, a homogeneous configuration of such a gauge field $A_y(t)$ leads to the ``electric field'' $E_y=\dot{A}_y$ where the dot being time derivative. Indeed, the inflaton field in \cite{Arkani-Hamed:2003xts,Arkani-Hamed:2003wrq} is a homogeneous Wilson line mode $\theta(t)=\oint dy A_y(t,y)=\oint dyA_y^{(0)}(t)=2\pi R A_y^{(0)}(t)$ where $R$ being the radius of the fifth dimension and $A^{(0)}_y$ is the zero mode of the gauge field along compact direction. Note however that it is also possible that the gauge potential/Wilson line being not an inflaton but a light moduli that start to move at some time in the history of the Universe. In this work, we do not address the origin of the electric field but consider its consequences. 

In this work, we investigate some consequences of the electric field along a compact space in 5D $U(1)$ gauge theory where one spatial direction is compactified to a circle ${\mathbb R}^{1,3}\times {\mathbb S}^1$. In particular, we consider time evolution of a massless complex scalar field charged under the $U(1)$ gauge theory. One immediately finds that the charged field is either accelerated or decelerated by the electric field as is the case in non-compact spacetime. From a 4D effective theory viewpoint, the change of momentum along the compact direction looks like a change of the KK mass towers, instead. Furthermore, a KK mode eventually becomes a ``zero mode'' when it is ``decelerated'' enough, and such a mode is non-perturbatively produced via the Sauter-Schwinger effect~\cite{Sauter:1931zz,Schwinger:1951nm}, which we call {\it KK Schwinger effect}.\footnote{In \cite{Friedmann:2002gx}, the authors considered similar processes to ours but their ``electric field'' originates from the geometry of the compact space and is rather similar to Unruh effect~\cite{Unruh:1976db} or Hawking radiation~\cite{Hawking:1975vcx}.} One of the most important properties of this effect is that as long as total deceleration is large enough KK particle and anti-particle pair can be created however small the electric field is. Thus, KK particles can be produced within physics at the energy scale below the KK scale. We should emphasize that such a phenomenon is first discussed in~\cite{Furuuchi:2015foh} but we would like to discuss the phenomena from a different view,  {\it the validity of 4D EFT within a time-dependent background}. We will investigate the phenomenon by explicitly quantizing the charged scalar field in the time-dependent background, which was not done in \cite{Furuuchi:2015foh}. We also briefly discuss the backreaction dynamics associated with KK particle production. Furthermore, there are several related works: In \cite{Qiu:2020gbl}, the authors consider the Schwinger effect within $(1+1)$D model where the 1D space is compactified to a circle, which is quite similar to our setup except the presence of 3-spaces in our case. They evaluate time-dependent particle numbers as we do in this work. In~\cite{Brown:2015kgj,Draper:2018lyw}, the Schwinger effect in thermal equilibrium is discussed where the imaginary time direction rather than a space direction is compactified, and the production rate is computed with the world line instanton method~\cite{Dunne:2005sx}. 

We should emphasize that the production of KK particles we discuss here is quite different from the swampland conjectures~\cite{Vafa:2005ui,Ooguri:2006in} (see e.g. \cite{vanBeest:2021lhn,Agmon:2022thq} for recent reviews) in the following sense: The distance conjecture~\cite{Ooguri:2006in} claims that a large field excursion of a field $\phi$ leads to the infinite towers of states being light as $M_n\propto e^{-\Delta\phi}$ where $\Delta\phi$ denotes the field excursion of $\phi$. On the other hand, in the model we discuss, what physically happens is the relabeling of the KK number due to the acceleration/deceleration by electric field, and there appear no infinite light KK modes but just a single light mode. Instead, the KK Schwinger effect we discuss here tells that naive truncation of KK modes at a given time cannot be justified when gauge fields along compact dimensions change in time, namely, 4D effective field theory description has to be carefully treated particularly in cosmological models. 

We also emphasize that our observation does not rely on the details of the compactification models since the mechanism itself is rather simple and general, although we discuss the simplest model in 5D. Therefore, similar consideration should be applied to any compactification models as long as they have candidates for electric fields along compact spaces and charged fields.

\section{Setup}
We consider a massless complex scalar field $\Phi$ coupled to U(1) gauge field in $(4+1)$D spacetime where the extra space is compactified to a circle $S^1$ $ds^2=\eta_{\mu\nu}dx^\mu dx^\nu+(2\pi R)^2dy^2$. We assume a periodic boundary condition $y\sim y+1$. The action of the complex scalar $\Phi$ is given by
\begin{align}
    S=&-\int d^5X D_M\bar{\Phi} D^M\Phi\nonumber\\
    =&-\int d^4xdy(2\pi R) (\partial_M-\ri qA_M)\bar\Phi g^{MN}(\partial_N+\ri qA_N)\Phi,
\end{align}
where $q$ denotes a charge of the complex scalar $\Phi$.
We assume a background electric field along the compact space by
\begin{align}
    A_y=\zeta(t)\to E_y=F_{0y}=\dot{\zeta},
\end{align}
where $A_y$ is a vector potential of the circle direction and $\zeta$ is a dimensionless real function of time. One can easily expand the complex scalar as
\begin{align}
    \Phi=\frac{1}{\sqrt{2\pi R}}\sum_{n\in \mathbb{Z}}\phi_{n}(x)e^{2\pi\ri n y},
\end{align}
which satisfies the above periodicity condition. One can easily perform KK decomposition of the system
\begin{align}
    S=-\sum_{n\in \mathbb Z}\int d^4x\left[\partial_\mu\bar{\phi}_n\partial^\mu\phi_n+\frac{1}{(2\pi R)^2}(n+q\zeta)^2|\phi_n|^2\right].
\end{align}
Thus, the KK modes acquire a time dependent mass. 
\begin{align}
    M_n^2(t)=\frac{1}{(2\pi R)^2}(n+q\zeta(t))^2.\label{effKKmass}
\end{align}
From a quantum gravity viewpoint, $U(1)$ charges are expected to be quantized~\cite{Banks:2010zn}, namely $q$ is an integer, and we will assume such a condition. Notice that $\zeta(t)\to \zeta(t)+1$ leads to $M_n^2(t)\to \frac{1}{2\pi R}(n+q+q\zeta(t))^2$ and relabeling of the KK number $\tilde{n}=n+q$ does not change the action, and in this sense, only the value of $q\zeta(t)-\lfloor q\zeta(t)\rfloor$ has the physical meaning in the above classical action as is clearly discussed in \cite{Furuuchi:2015foh}.

Let us discuss the quantization of $\phi_n$. We formally expand a quantum operator $\hat\phi_n$ as
\begin{align}
   \hat\phi_n(t,\bm x)=&\int \frac{d^3\bm k}{(2\pi)^{\frac32}}\left[\hat{a}_{n,\bm k}f_{n,k}(t)e^{+\ri\bm k\cdot \bm x}+\hat{b}_{n,\bm k}^\dagger \bar{f}_{n,k}(t)e^{-\ri \bm k\cdot \bm x}\right],\\
    \hat{\bar\phi}_n(t,\bm x)=&\int \frac{d^3\bm k}{(2\pi)^{\frac32}}\left[\hat{b}_{n,\bm k}f_{n,k}(t)e^{+\ri\bm k\cdot \bm x}+\hat{a}_{n,\bm k}^\dagger \bar{f}_{n,k}(t)e^{-\ri \bm k\cdot \bm x}\right],
\end{align}
where $\hat{a}_{n,\bm k},\hat{b}_{n,\bm k}^\dagger$ are creation and annihilation operators satisfying $[\hat{a}_{n,\bm k},\hat{a}^{\dagger}_{n',\bm k'}]=\delta_{n,n'}\delta^3(\bm k-\bm k')$ and $[\hat{b}_{n,\bm k},\hat{b}^{\dagger}_{n',\bm k'}]=\delta_{n,n'}\delta^3(\bm k-\bm k')$, respectively. The mode function $f_{n,k}$ follows the equation of motion of $\phi_n$,
\begin{align}
    \ddot{f}_{n,k}(t)+(k^2+M_n^2(t))f_{n,k}(t)=0.
\end{align}
Assuming $\zeta(t)\to 0$ for $t<t_0$ and we take an adiabatic initial condition $f_{n,k}(t)\to \frac{1}{\sqrt{2\omega_{n,k}(t)}}\exp(-\ri\int^t\omega_{n,k}(t')dt')$, \footnote{The lower limit of the integral in the exponent is arbitrary, but we consider it to be some time $t<t_0$.} where $\omega^2_{n,k}(t)\equiv k^2+M_{n}^2(t)$. Then, one can define an adiabatic vacuum state $|0\rangle_{\rm in}$ that satisfies $\hat{a}_{n,\bm k}|0\rangle_{\rm in}=\hat{b}_{n,\bm k}|0\rangle_{\rm in}=0$, which corresponds to the initial vacuum state. In general, the construction of an adiabatic state is not unique and here we take the lowest order adiabatic solution as\footnote{Thus defined adiabatic vacuum is not a unique choice as only asymptotic condition is imposed. Changing the adiabatic expansion order gives different behavior of time-dependent particle number densities~\cite{Dabrowski:2014ica,Dabrowski:2016tsx,Yamada:2021kqw}. However, when the background field asymptotically decays, the particle number becomes independent of the choice of the adiabatic expansion order.}
\begin{align}
    f_{n,k}(t)=\frac{1}{\sqrt{2\omega_{n,k}(t)}}\Biggl[&\alpha_{n,k}(t)\exp\left(-\ri\int^t\omega_{n,k}(t')dt'\right)\nonumber\\
    +&\beta_{n,k}(t)\exp\left(+\ri\int^t\omega_{n,k}(t')dt'\right)\Biggr],
\end{align}
where the auxiliary functions $\alpha_{n,k}(t)$ and $\beta_{n,k}(t)$ satisfy
\begin{align}
&\dot{\alpha}_{n,k}=\frac{\dot{\omega}_{n,k}}{2\omega_{n,k}}\beta_{n,k}e^{+2\ri \int^t \omega_{n,k}(t')dt'},\quad
    \dot{\beta}_{n,k}=\frac{\dot{\omega}_{n,k}}{2\omega_{n,k}}\alpha_{n,k}e^{-2\ri \int^t\omega_{n,k}(t')dt'},\label{alphabetaeq}\\
    &|\alpha_{n,k}(t)|^2-|\beta_{n,k}(t)|^2=1,
\end{align}
and the adiabatic initial condition is now rewritten as $\alpha_{n,k}(t)\to 1$, $\beta_{n,k}(t)\to 0$ for $t<t_0$. Using these quantities, one is able to evaluate the amount of the vacuum pair production by
\begin{align}
    \langle \hat{N}_{n,\bm k}(t)\rangle=|\beta_{n,k}(t)|^2\label{pn},
\end{align}
where $\langle \hat{N}_{n,\bm k}(t)\rangle$ is the number density of (anti-)particles in the phase space. Note that thus defined particle density is ambiguous within a finite time interval, but becomes definite only when $\zeta\to{\rm const.}$ asymptotically. 

Despite difficulty of finding analytic solutions to \eqref{alphabetaeq}, we are able to use numerical integration to solve them, from which we are able to evaluate \eqref{pn} and in the following section, we show some examples of numerical solutions.

\section{KK Schwinger effect}
We discuss the KK Schwinger effect due to the electric field along the compact direction. One of the most important properties of the electric field in compact spaces is that it ``accelerates'' the KK momentum as the case of electric fields in non-compact 4D spacetime case. From a 4D effective theory viewpoint, a KK mode at a given time becomes a zero mode at a later time due to the change of the time-dependent mass. Therefore, production of KK particles may not cost much energy than usually expected. 

Let us first discuss the energy scales of the electric field. The action of the U(1) gauge field is
\begin{align}
    S_{\rm vec}=-\frac{1}{4g^2}\int d^4x\int_{0}^1dyF^{MN}F_{MN}\to\frac{1}{2g^2}\int d^4x\frac{1}{(2\pi R)}\dot{\zeta}^2,
\end{align}
from which one finds the energy density of the electric field in 4D effective theory to be
\begin{align}
    \rho_E= \frac{\dot{\zeta}^2}{4\pi Rg^2}\equiv\frac12\dot{\zeta}_c,
\end{align}
where $\zeta_c=(2\pi Rg^2)^{-1/2}\zeta$ is a canonically normalized scalar field in 4D theory. Hereafter, we will take $1/g^2=M_{\rm KK}=1/(2\pi R)$ for simplicity. In the following, we assume that $|\dot{\zeta}_c|\ll M_{\rm KK}^2$ from which one may naively expect that the 4D effective theory is valid. On the other hand, the cost of KK particle production is determined also by the ``total change of momentum'' $\Delta\zeta$ as is clear from the effective KK mass formula~\eqref{effKKmass}. Indeed, as we see below, the KK particles are created when $M_n^2(t)\to 0$. Therefore, whether or not $|q\Delta\zeta|>1$ or in the canonical language, $|\Delta\zeta_c|>M_{\rm KK}$ is crucial for the KK particle production. Let us emphasize again that the condition $|\Delta\zeta_c|>M_{\rm KK}$ is not that for KK particles being light but that for a ``KK particle'' at a given time becomes a ``zero mode'' at a different time. Therefore, despite resemblance to the distance conjecture condition~\cite{Ooguri:2006in}, it is not really the same one since the towers of states do not become light. 

As an illustration, we consider examples of $\zeta(t)$ and evaluate the amount of KK particle produced from the initial vacuum.
We here consider two toy models, (i)~$\zeta_c(t)=Et\Theta(t)\Theta(t_f-t)$ and (ii) $\zeta_c(t)=\frac{E}{m}e^{-\gamma t}\cos(m t)$ where $E,\gamma$ and $m$  are real parameters, and $\Theta(t)$ is the Heaviside step function, $\Theta(t)=1$ for $t\geq0$ and $\Theta(t)=0$ for $t\leq0$, and we will take $t_0=0$. These field configurations are chosen just for illustrating purposes but we discuss possible origins of the electric field in Sec.~\ref{origin}.
\subsection{Model (i)}\label{model1}
\begin{figure}[htbp]
    \centering
\includegraphics[keepaspectratio, scale=0.6]{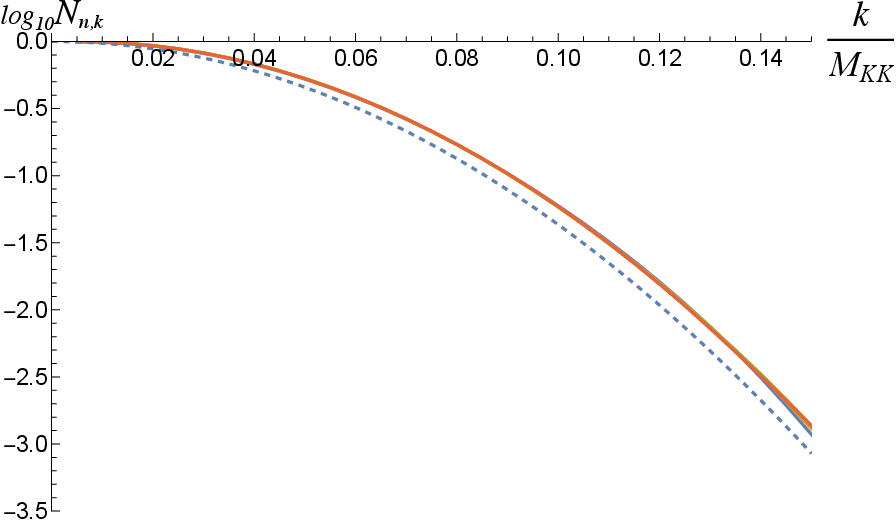}
    \caption{The KK particle number density for $n=-1,-2,-3,-4$ (solid lines) and the approximate formula~\eqref{model1Nformula} (dashed line) at $t=6\times10^{2}$ for $qE=10^{-2}$ in the unit $M_{\rm KK}=1$ (Model(i)). As expected, the number density is independent of the KK number $n$ and all solid lines are almost indistinguishable.}
    \label{fig:model11}
    \end{figure}

\begin{figure}[htbp]
    \centering
\includegraphics[keepaspectratio, scale=0.8]{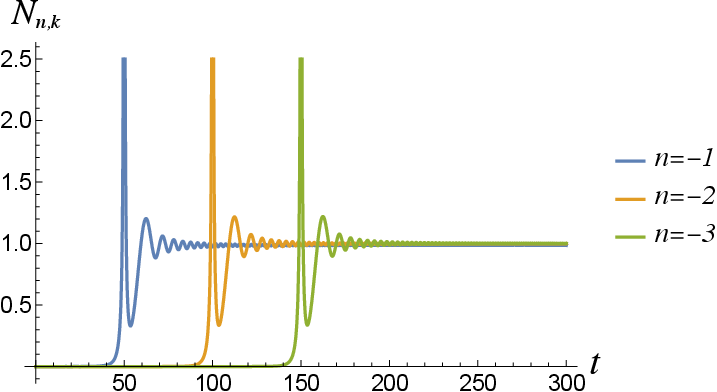}
    \caption{The KK particle number density for $n=-1,-2,-3$ as functions of time $t$ for $qE=10^{-2}$ in the unit $M_{\rm KK}=1$ (Model(i)). The particle numeber becomes nonzero around the time at which the effective KK mass vanishes for all modes. }
    \label{fig:model12}
    \end{figure}

For the model (i), the constant electric field exists only for $0\leq t\leq t_f$. Notice that the effective KK mass~\eqref{effKKmass} becomes zero when $t_{n}=-\frac{n}{qE}$, and therefore, we expect the ``decelerated'' mode that crosses $M_n^2=0$ would be produced efficiently. Assuming $qE>0$, we are able to make an analytic estimate of the number density e.g. by using the well known formula for Schwinger pair production with a constant electric field as
\begin{align}
    \langle \hat{N}_{n,k}\rangle=\exp \left(-\frac{\pi k^2}{qE}\right) \qquad \text{for }n\leq -\lfloor qEt_f\rfloor.\label{model1Nformula}
\end{align}
Notice that there is no KK number dependence since the pair production occurs when the KK momentum is decelerated to zero. Therefore, if the electric fields are turned on for a sufficiently long term, multiple KK particles can be produced however small $E$ is. We show numerical examples of the produced number density in Fig.~\ref{fig:model11} and its time dependence in Fig.~\ref{fig:model12}. In the examples, we have taken $E=0.01M_{\rm KK}^2$ and therefore, the electric field strength is smaller than the naive cut-off scale $M_{\rm KK}$. Nevertheless, as expected, each KK mode can be produced.


\subsection{Model (ii)}\label{model2}
\begin{figure}[htbp]
    \centering
\includegraphics[keepaspectratio, scale=0.8]{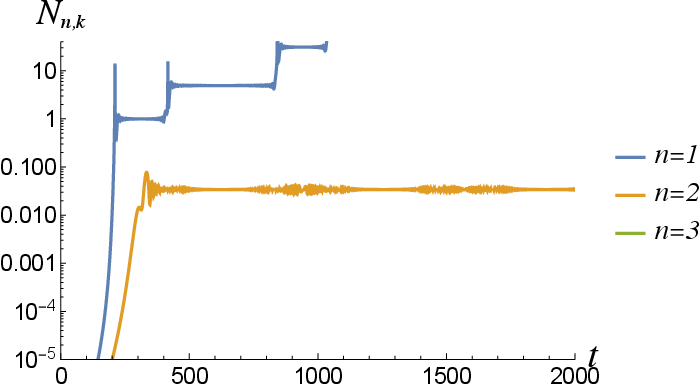}
    \caption{The KK particle number density for $n=1,2,3$ as functions of time $t$ for $qE=2\times 10^{-2},m=10^{-2}, \gamma=10^{-4}, k=10^{-3}$ in the unit $M_{\rm KK}=1$ (Model(ii)). The resonance leads to large number densities for $N_{n,k}$ for $n=1$ but since $|\Delta\zeta_c|<2M_{\rm KK}$ the KK mode $n=2$ is less than $n=1$ and no resonance effect. The produced number of $n=3$ mode is negligibly small since the KK mass never becomes small.  }
    \label{fig:model21}
    \end{figure}

For the model (ii) where the modulus experiences a damped oscillation, the amplitude of the oscillation determines whether KK modes are produced or not. Assuming $qE>0$, the first minimum of the effective mass for $n>0$ appears around $t=\pi/m$ and takes the value $M_{n}^2(\pi/m)=(n-qE/m e^{-\pi\gamma/m})^2$. Hence, the modes with their KK number satisfying $\lfloor qE/m e^{-\pi\gamma/m}\rfloor>n>0$ can be produced. In the oscillatory case, the burst of particle production may occur due to the resonance like the case of the preheating after inflation~\cite{Kofman:1997yn}. Indeed, as shown in Fig.~\ref{fig:model21}, $n=-1$ mode experiences resonant particle production. As we expected, for the modes $n>|\Delta\zeta_c|/M_{\rm KK}$ cannot be created (see $n=-3$ case of Fig.~\ref{fig:model21}). Thus, repeated KK Schwinger effects may cause large number of KK particle production despite the smallness of the electric field. Note also that this is a consequence of the background field approximation, and the backreaction would stop the violent KK particle production as we will discuss next. We omit the analytical estimate of produced particle number density, it can be estimated by analyzing Stokes phenomenon of the mode equations. See~\cite{Dumlu:2010ua,Dabrowski:2014ica,Dabrowski:2016tsx,Taya:2020dco,Yamada:2021kqw} for details.
\subsection{Why does the KK Schwinger effect matter?}\label{origin}
We have seen that electric field with energy much below KK scale can create KK particles. In this subsection, we discuss 1. possible origins of electric field, 2. implications of the KK Schwinger effect to the 4D effective theory description and 3. the fate of KK particles created from vacuum.

First, we have assumed electric field, which can be justified only when $\zeta$ is light enough. The lightness at tree level is ensured by gauge invariance of the original action. However, one notice that since the gauge field necessarily couples to charged particles, there appear quantum corrections such as Coleman-Weinberg potential of the gauge potential. Such a correction may be suppressed by supersymmetry. In particular, higher-dimensional supersymmetry corresponds to extended supersymmetry in 4D and the quantum corrections can further be suppressed than one expects. In such a situation, the gauge potential can be light enough to experience large field excursion. Note that if electric field is turned on, the time-translation symmetry is violated and therefore supersymmety would be spontaneously broken. Interestingly, in such a case, the quantum corrections should be proportional to $E_y=\dot\zeta$ as the constant gauge potential cannot break supersymmetry.\footnote{We do not assume non-trivial boundary condition such as Scherk-Schwarz compactification~\cite{Scherk:1979zr}, which does break supersymmetry.} Therefore, the associated corrections would not give ``potential'' of $\zeta$ but derivative corrections. We will investigate it elsewhere. Other possible potential of $\zeta$ may originate from non-perturbative corrections, but it can be small in general. Therefore, if the potential is protected against quantum corrections, the light scalar $\zeta$ may vary over a large range of its field space. Another possibility would be an accidental flatness of the potential for $\zeta$. For instance,  when multiple bosonic and fermionic particles having different bulk masses and charges contribute to the quantum corrections to $\zeta$ effective potential, it is possible that the effective potential becomes flat over the range beyond KK scale. If effective potential is sufficiently flat, inflation may take place by the quantum effective potential. In such a case, particle production during inflation must be taken account of.

Secondly, let us discuss what is the lesson from the KK Schwinger effect. In constructing 4D cosmological models from higher-dimensional theory, one often truncates KK modes and construct a 4D effective Lagrangian of zero modes, from which one discusses the dynamics of effective 4D universe. However, as discusses here, time evolution of gauge potential along compact directions may change the definition of ``zero modes''. We should note however that unless particle production is taken account of, the change of the definition of ``zero modes'' cannot be physical at all. Indeed, as discussed below \eqref{effKKmass}, shift of field value of $\zeta$ is unphysical, and can simply be absorbed into the redefinition of the KK label $n$. Note that for instance, the quantum corrections associated with KK particles are not affected by the relabeling since the KK sum is taken over all $n$. However, if KK particles are materialized, the truncation of KK modes done at the beginning is no longer justified. Of course, this is not a physical problem, but a technical one: We should have left the KK modes that can be created in the later time.  In this sense, the ``4D effective theory'' given by integrating out KK modes becomes invalid when the KK Schwinger effect takes place. In other words, when the notion of zero modes varies over time, we should use a refined notion of 4D effective theory that include ``KK modes'' that may become ``zero mode'' during the time evolution.

Thirdly, we briefly discuss the fate of the KK particles: If the created KK particles suddenly decay to lighter particles, there would be no consequences of KK particle production except possible imprints on cosmic microwave backgrounds. Notice however that the translational invariance of compact space implies the KK momentum to be a conserved quantity. Therefore, the number density of the lightest KK particles must be coannihilation of pairs. In such a case, the number density of the lightest KK particle would not decay exponentially as the coannihilation probability becomes small as the particle number density is reduced. If they are relatively long lived, the KK particles behave as heavy moduli particles, which affect the thermal history of the universe. Furthermore, if $\zeta$ becomes inflaton varying over the scale greater than KK scale (possibly by the scenario mentioned above), the particle production during inflation may leave imprints on the spectrum of cosmic microwave background~\cite{Green:2009ds,Flauger:2016idt}. We will leave the physics of the produced KK particle, but emphasize that thus-produced KK particles may affect low energy physics even if they do not remain in the universe today.

\section{The backreaction from KK particle production}\label{backreaction}
As shown in the previous section, KK particles are produced even though the energy of the electric field is smaller than the KK scale, which implies that 4D EFT derived by truncation of KK modes at the initial time cannot be valid. In particular, the resonant particle production may produce enormous numbers of KK particles.

\begin{figure}[htbp]
    \centering
\includegraphics[keepaspectratio, scale=0.8]{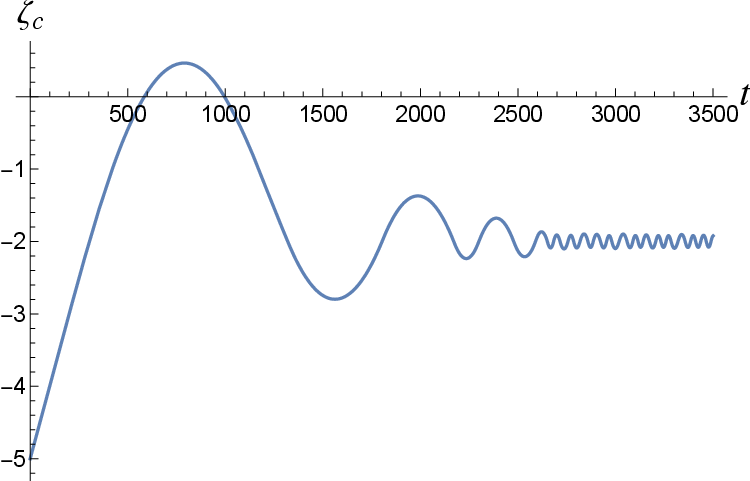}
    \caption{The numerical solution of coupled equations~\eqref{alphabetaeq} and~\eqref{zetaEOM}. The unit is taken to be KK scale $M_{\rm KK}=1$ and the initial velocity $\dot{\zeta}_c(0)=0.01$, $\zeta_c(0)=-5$. The cut-off momentum is $k_{\rm max}=\sqrt{|q\dot{\zeta}_c(0)|}$, the 3-momentum lattice spacing $N=8$ and the charge $q=1$.}
    \label{fig:dynamics}
    \end{figure}

However, we argue that such violent KK particle production may not continue due to the backreaction to the electric field: Physically speaking, the produced KK pairs induce an electric current along the compact direction, which shields the electric force, and eventually effective electric force would become zero after which the pair production stops. Indeed, the equation of motion of $\zeta$ is\footnote{This is essentially the E.O.M. of $A_y$ and can be derived from 5D action with an appropriate gauge fixing.}
\begin{align}
    \ddot{\zeta}_c+\sum_{n\in \mathbb Z}2q\left(nM_{\rm KK}+q\zeta_c\right)\langle |\hat{\phi}_n|^2\rangle_{\rm ren}=0\label{zetaEOM}
\end{align}
where $\langle \cdots\rangle_{\rm ren}$ denotes the renormalized expectation value. The second term on the left-hand-side of the above equation is nothing but the effect of the electric current. The un-renormalized expectation value $\langle |\hat{\phi}_n|^2\rangle$ is 
\begin{align}
    \langle |\hat{\phi}_n|^2\rangle=&\int \frac{d^3k}{(2\pi)^3}\frac{1}{2\omega_{k,n}(t)}|f_{n,k}(t)|^2\nonumber\\
    =&\int \frac{d^3k}{(2\pi)^3}\frac{1}{2\omega_{k,n}(t)}\left[|\alpha_{k,n}(t)|^2+|\beta_{k,n}(t)|^2+2{\rm Re}\left(\alpha_{k,n}\bar{\beta}_{k,n}e^{-2\ri\int^{t} dt'\omega_{k,n}(t')}\right)\right]\nonumber\\
    =&\int \frac{d^3k}{(2\pi)^3}\frac{1}{2\omega_{k,n}(t)}\left[1+2|\beta_{k,n}(t)|^2+2{\rm Re}\left(\alpha_{k,n}\bar{\beta}_{k,n}e^{-2\ri\int^{t'} dt'\omega_{k,n}(t)}\right)\right],
\end{align}
where in the third equality we have used $|\alpha_{k,n}(t)|^2-|\beta_{k,n}(t)|^2=1$. The constant in the square bracket causes UV divergence and we simply remove it e.g. by assuming underlying supersymmetry, which cancel such divergence between bosons and fermions as is assumed in the model of~\cite{Kofman:2004yc}.\footnote{As we mentioned in Sec.~\ref{origin}, there may appear corrections proportional to $\dot\zeta$. However, we omitt it here and leave its effect for future work.} Then we define the renormalized two point function to be
\begin{align}
    \langle |\hat{\phi}_n|^2\rangle_{\rm ren}=&\int \frac{d^3k}{(2\pi)^3}\frac{1}{\omega_{k,n}(t)}\left[|\beta_{k,n}(t)|^2+{\rm Re}\left(\alpha_{k,n}\bar{\beta}_{k,n}e^{-2\ri\int^{t'} dt'\omega_{k,n}(t)}\right)\right]\nonumber\\
    \approx& \int \frac{d^3k}{(2\pi)^3}\frac{1}{\omega_{k,n}(t)}\langle \hat{N}_{k,n}(t)\rangle,
\end{align}
where we have omitted the oscillatory term since it oscillates very fast and the average would be zero and used ~\eqref{pn}. Note that the effect of particle production does not cancel between bosons and fermions even if supersymmetry exists. We assume that this is the leading order correction to the effective potential of $\zeta$. In such a case, the gauge potential would acquire a constant velocity, namely constant electric field would be realized. In general, it is not easy to solve the coupled equations since quantum field has infinite number of $(k,n)$ modes (even if $k$ is discretized). Nevertheless, we examine the numerical solution to the coupled differential equations by taking account of a few modes as is done in \cite{Kofman:2004yc,Tanji:2008ku,Kikuchi:2023uqo}.\footnote{In \cite{Tanji:2008ku,Kitamoto:2020tjm}, the backreaction from pair creation is analytically discussed, but it requires some approximations and the correct behavior would in general require numerical analysis.}  Here we discretize the momentum $k\to k_j=jk_{\rm max}/N$ ($j=0,\cdots,N$, $N\in {\mathbb N}$) with the cut-off momentum $k_{\rm max}$ and approximate
\begin{align}
\langle |\hat{\phi}_n|^2\rangle_{\rm ren}\to \frac{1}{2\pi^2}\sum_{j=0}^{N}\left(\frac{k_{\rm max}}{N}\right)^3\frac{j^2\langle \hat{N}_{k_j,n}(t)\rangle}{\omega_{k_j,n}}.
\end{align}
We take $N=8$ as lattice spacing and $k_{\rm max}=\sqrt{|q\dot{\zeta}_c(0)|}$ as the cut-off 3-momentum.\footnote{This is reasonable choice as seen from \eqref{model1Nformula} since the production rate is suppressed when 3-momentum exceeds the electric field strength.} We consider only the contributions from $n=-3,\cdots, 3$ since not all the KK modes are produced. With these approximation and the parameter set, we have numerically solved coupled equation of motion of each $(n,k_j)$ mode~\eqref{alphabetaeq} and of $\zeta(t)$~\eqref{zetaEOM} as shown in Fig.~\ref{fig:dynamics}. The first particle production event is described by the model (i) in Sec.~\ref{model1}, but backreaction eventually traps $\zeta_c$ to some point in the field space. Notice that $\zeta_c$ in this case experiences the decaying oscillation similar to the model (ii) in Sec.~\ref{model2}. Despite rough approximation, we find the backreaction to stop KK particle production eventually.\footnote{In cosmological background, one has to take account of the effect of expanding universe, which behaves as friction like force.} Improving the number of lattice spacing and taking more KK modes would change the result quantitatively, but the qualitative behavior would not change, which however requires more computational costs. 


Strictly speaking, the improved 4D effective theory is not completely correct since the production of the KK modes implies that there appear the momentum current of a compact space namely $T_{y0}$ and $T_{yy}$ where $T_{MN}$ is the energy momentum tensor. Although we have neglected the effect to the background geometry, one also has to consider the effect on the compactification, or more specifically, the effect on the radius stabilization of the compact space through 5D Einstein equation. Such effects may become important if the radion field of the compact space is light but we will not investigate its detail here.

\section{Summary}
We have investigated the role of electric fields in compact directions and find that the acceleration or deceleration of KK momenta may cause non-perturbative KK particle production via Schwinger effect, which cannot be seen within perturbative processes. One of the most remarkable properties of this phenomenon is that the KK particle production can occur when the integrated deceleration $\Delta\zeta$ is large enough even though the electric field strength $\dot\zeta$ is smaller than the KK scale $M_{\rm KK}=\frac{1}{2\pi R}$. Such an effect would be independent of the details of the compactification. Let us give a few comments on the generality of the KK Schwinger effect. If zero modes of gauge field component along compact directions, as long as they are light, it is possible for KK particles to be produced, and the number of compact dimensions is not important and can be higher than one. In particular, critical superstring theory predicts six extra dimensions. Orbifold projection removes some of such gauge fields but it is in general hard to remove all the gauge potentials along compact spaces in realistic compactification. As long as there exist such fields and they time-evolve over a large field range beyond compactification scales, KK particles would be produced via KK Schwinger effects, which may spoil a naive 4D effective description given by the truncation of all the KK modes at initial time. It is also possible that the charged matter has a bulk mass term, which is not assumed in this work. If a charged field has a bulk mass term heavier than the electric field scale, its zero mode as well as KK modes would not be produced much, which is another way to avoid the KK particle production.

As we argued, the KK pair production backreacts to the motion of the modulus, and eventually the electric field would be turned off as explicitly shown in Sec.~\ref{backreaction}. Such a mechanism would also have some applications to the physics of open string moduli of D-brane systems in string theory. Indeed, the KK Schwinger effect is related to the modulus trapping mechanism proposed in~\cite{Kofman:2004yc}. D$p$-brane action can be thought of dimensional reduction of D$9$-brane with the identification $A_i\to X^i$ $(i=9-p)$ where $A_i$ is the gauge field associated with the open string attached to D$9$-brane while $X^i$ being the position moduli of D$p$-brane. Therefore, KK Schwinger effects would also be relevant e.g. to brane inflation~\cite{Dvali:1998pa,Kachru:2003sx} where position moduli of D$p$-branes play the role of inflaton fields. The backreaction of pair production during inflation may leave interesting observational consequences~\cite{Green:2009ds,Flauger:2016idt}. It would be worth revisiting some inflationary models based on the higher-dimensional gauge fields with the renewed interest from the KK Schwinger effect viewpoint. In particular, the idea of the trapped inflation~\cite{Green:2009ds} may be realized by including KK mode production, and we leave such an interesting possibility for future work.

One of the lessons from our observation is that naively truncated heavy modes may become light and can be materialized as real particles even within low energy physics and the validity of the naive effective theory can fail. In this sense, one has to carefully consider the dynamics in the seemingly unobservable compact spaces.

\section*{Acknowledgements}
YY would like to thank Hiroyuki Abe for discussions on related issues and is supported by Waseda University Grant for Special Research Projects (Project number: 2023C-584).

\end{document}